\documentclass[conference]{IEEEtran}
\IEEEoverridecommandlockouts
\usepackage{cite}
\usepackage{amsmath,amssymb,amsfonts}
\usepackage{textcomp}
\usepackage{amsthm}
\usepackage{graphicx}
\usepackage{multirow}
\usepackage{textcomp}
\usepackage{comment}
\usepackage[font=footnotesize]{caption}
\usepackage[font=footnotesize]{subcaption}
\usepackage{xcolor}
\usepackage[noend]{algpseudocode}
\usepackage{algorithm}
\usepackage{url}
\def\BibTeX{{\rm B\kern-.05em{\sc i\kern-.025em b}\kern-.08em
    T\kern-.1667em\lower.7ex\hbox{E}\kern-.125emX}}

\newtheorem{theorem}{\textbf{Theorem}} 
\newtheorem{lemma}{\textbf{Lemma}}

\newcounter{lemtheorem}
\newtheorem{definition}{\textbf{Definition}}

\DeclareMathOperator*{\argmin}{\text{arg min}\,}
\DeclareMathOperator*{\argmax}{\text{arg max}\,}

\begin{document}


\title{Congestion Mitigation in Vehicular Traffic Networks with Multiple Operational Modalities\\}

\author{\IEEEauthorblockN{Doris E. M. Brown and Sajal K. Das}
\vspace{0.5ex}
\IEEEauthorblockA{
Department of Computer Science 
\\ 
Missouri University of Science and Technology 
\\ 
Rolla, MO 65409, USA 
\\ Email: \{deby3f, sdas\}@mst.edu}
}

\maketitle

\begin{abstract}
Modern commercial ground vehicles are increasingly equipped with multiple operational modalities, including human driving, advanced driver assistance, remote tele-operation, and full autonomy. These modalities often rely on heterogeneous sensing infrastructures and distinct routing algorithms, which can yield misaligned perceptions of the traffic environment and competing route preferences. While such technologies  accelerate the transition toward increasingly intelligent transportation networks, their current deployment fails to avoid challenges associated with selfish routing behavior, in which drivers or automated agents prioritize individually optimal routes at the expense of network-wide congestion mitigation. Promising existing traffic flow management strategies can address leader-follower dynamics in traffic routing problems but are not designed to account for vehicles capable of dynamically switching between multiple operational modes, each with potentially misaligned objectives and beliefs. This paper models the interaction between a vehicle control arbitration system and a multi-modal vehicle as a repeated single-leader, multiple follower Stackelberg game with asymmetric information. To address the intractability of computing an exact solution in this setting, we propose a {\em Trust-Aware Control Trading Strategy} (TACTS) utilizing a regret matching-based algorithm to adaptively update the arbitration system’s mixed strategy over sequential, dynamic routing decisions. Theoretical results provide bounds on the realized total network travel time under TACTS algorithm relative to the system-optimal total network travel time. Experimental results of simulations between the system and a vehicle in several real-world traffic networks under various different congestion levels demonstrate that TACTS consistently reduces network-wide congestion and generally outperforms alternative routing and control-allocation strategies, particularly under high congestion and heavy induced vehicle flows.
\end{abstract}


\vspace{-0.1in}
\section{Introduction}
\vspace{-0.05in}
Traffic congestion has been an increasingly prevalent problem in urban vehicular transportation systems worldwide, incurring significant economic, environmental, and social costs. A wide range of strategies have been proposed to mitigate congestion, from adaptive traffic signal control to incentive mechanisms and cooperative routing algorithms. However, most existing approaches rely on limiting or unrealistic assumptions about driver behavior and control paradigms by typically modeling vehicles as either fully human-driven, fully autonomous, or a fixed mixture of the two, for which congestion mitigation solutions leverage modality-specific assumptions such as perfect compliance from autonomous vehicles or fixed behavioral rules from human drivers. In reality, modern transportation networks already contain vehicles equipped with multiple operational modalities, such as human driving, advanced driver-assistance systems, remote tele-operation, and varying levels of autonomy that can be dynamically shifted during a single routing interaction through a traffic network. Furthermore, each operational modality may exhibit individual beliefs about the traffic network that dictate its routing decisions, resulting from different sensing infrastructure or misaligned motives. This evolving capability creates traffic dynamics that fall outside the scope of many existing mitigation approaches, limiting their practical effectiveness in realistic modern ground vehicle traffic networks in an intelligent transportation system.

Many existing works on congestion mitigation involve traffic routing strategies that assume increasing vehicle autonomy will enable system-optimal coordination of all or a subset of vehicles in the network \cite{bang2022combined, mostafizi2021decentralized, pei2023self, kashmiri2024routing}. However, the reality of current multi-modal vehicle behavior challenges this premise. Empirical observations of vehicles equipped with full self-driving capabilities suggest that, in practice, these systems often replicate the same selfish routing preferences observed in human drivers, often choosing routes that optimize individual travel time, safety, efficiency, and other metrics, rather than paths that optimize the total network metrics \cite{tesla2025model3_navigation, waymo_user_route_selection_2023}. This behavior is further reinforced by evidence that human users prefer routing choices aligned with their own comfort or driving style, even when they deviate from system-optimal solutions \cite{craig2021should}. Thus, even as vehicles transition from Level 0 to Level 5 autonomy \cite{mobilus2021sae}, the persistence of human-centric design objectives and routing strategies implies that emerging modalities may continue to exhibit self-interested behavior. This undermines existing works that assume autonomous vehicles will act in coordination with centralized traffic objectives, thus limiting the relevance of these models in increasingly heterogeneous, modality-flexible traffic networks.

\begin{figure}[!t]
\vspace{-0.1in}
\centering
\includegraphics[width=0.48\textwidth]{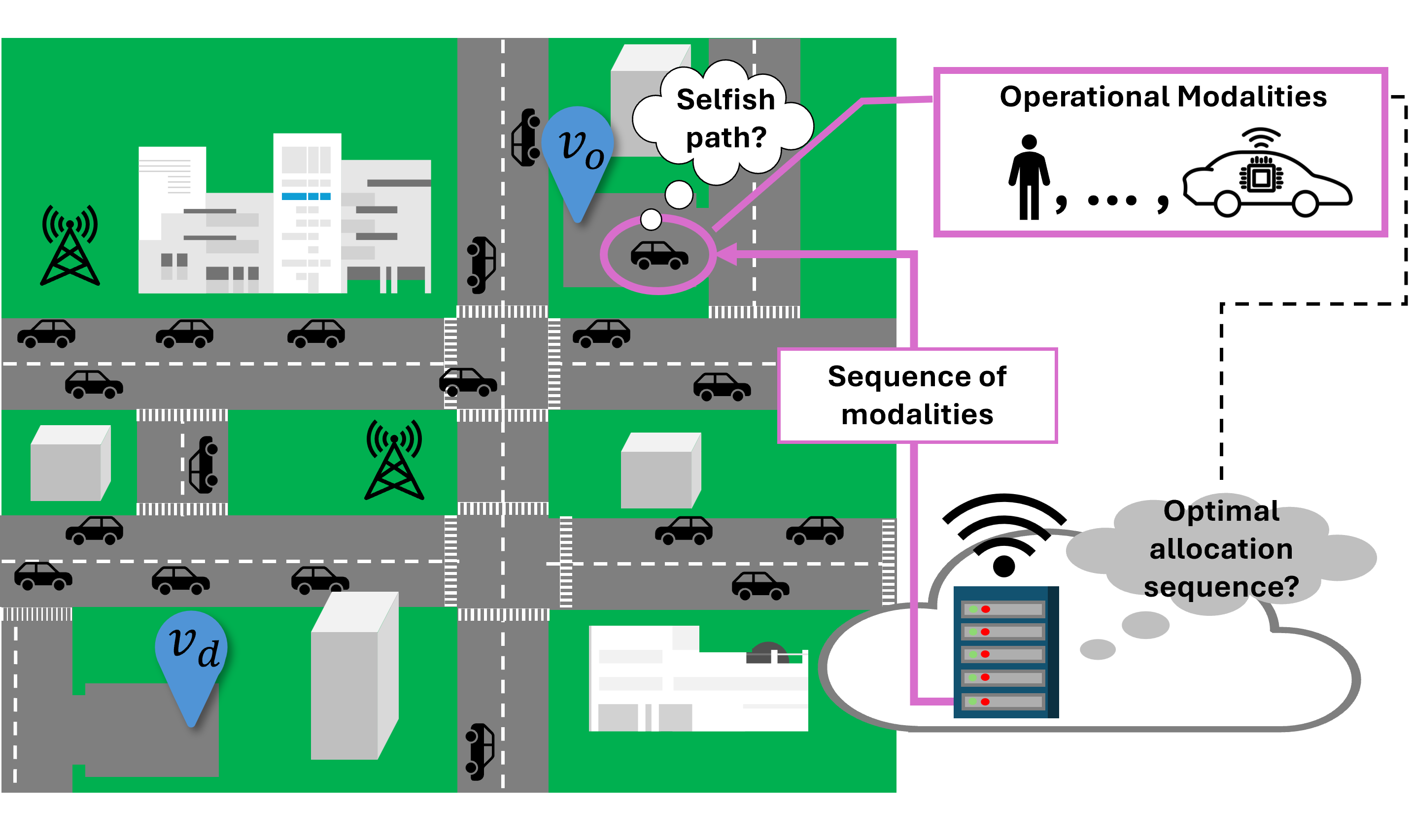}
\vspace{-0.15in}
\caption{An example of system-vehicle interaction in which a vehicle seeks sequence of operational modalities yielding selfish path from origin $v_o$ to destination $v_d$.}
\label{Img: Traffic Diagram}
\vspace{-0.25in}
\end{figure}

To address these significant challenges, this paper models the interaction between a central traffic arbitration system and a vehicle with multiple operational modalities, as a repeated Stackelberg game with asymmetric information. In this setting, the arbitration system seeks to minimize the total network travel time by allocating control of the vehicle to one of its operational modalities at each routing decision point, while the vehicle’s active modality acts to selfishly optimize its own travel objective of minimizing individual vehicle travel time. An example of this interaction in a traffic network is depicted in Fig. \ref{Img: Traffic Diagram}. 

The problem of determining an optimal traded control allocation strategy is inherently complex, given that the arbitration system must make control allocation decisions without full observability beliefs of each operational modality that govern their individual decision-making. To design a tractable and practically implementable solution, a {\em Trust-Aware Control Trading Strategy} (TACTS) is proposed, which uses a regret-matching approach to adaptively update the arbitration system’s mixed strategy over sequential routing decisions. The developed TACTS algorithm enables the arbitration system to anticipate the vehicle’s preferred actions under different modalities. This allows the vehicle to probabilistically prioritize trading vehicle control to operational modalities which have historically made decisions more closely aligned with those that would lead to a system-optimal routing outcome. 

\vspace{-0.1in}
\subsection{Novel Contributions of This Paper}
\vspace{-0.05in}
To the best of the authors’ knowledge, this work is the first to address congestion mitigation for vehicles with multiple operational modalities capable of dynamically switching operational modes, in the presence of misaligned beliefs and objectives between the vehicle and a vehicle control arbitration system. 
The primary contributions of this work are as follows.
\begin{enumerate}
    \item The routing interaction between a vehicle control arbitration system and a set of operational modalities is modeled as a \textbf{repeated Bayesian Stackelberg game}, where the system aims to construct a traded control strategy that minimizes the total travel time of the network and each operational modality aims to minimize the individual travel time of the vehicle.
    \item A \textbf{novel, trust-aware regret minimization-based control trading strategy} is developed to compute a  traded control strategy to mitigate the impact of the vehicle's routing decisions on the total network travel time.
    \item \textbf{Theoretical results} are derived for the performance ratio of the developed algorithm compared to the optimal total network travel time achievable through a traded control strategy employed by the system.
    \item \textbf{Simulation results} utilizing simulated traffic data on real traffic networks are presented. They demonstrate the resulting total network travel time, vehicle travel time, and execution time of the developed algorithm in comparison with state-of-the-art arbitration system traded control strategies and trust-aware routing strategies.
\end{enumerate}

The paper is organized as follows: Section \ref{Sec: Related Work} gives an overview of related work on traffic congestion mitigation. Section \ref{Sec: Problem Formulation} formulates the problem between the arbitration system and the set of operational modalities with misaligned motives. Section \ref{Sec: Proposed Methodology} details properties of a system-optimal traded control strategy and proposes TACTS, a trust-aware regret matching-based traded control system strategy to mitigate network travel time. Section \ref{Sec: Theoretical Results} analyzes the proposed algorithm, and Section \ref{Sect: Performance Evaluation} presents and describes experimental results.
Section \ref{Sect: Conclusion} concludes the paper.

\section{Related Work} \label{Sec: Related Work}
\vspace{-0.05in}
Traffic congestion in urban networks remains a persistent problem despite decades of research and numerous mitigation strategies. Proposed solutions generally fall into three categories: (a) managing and influencing travel modalities, such as expanding public transit, (b) increasing infrastructure capacity, such as adding lanes or new roads, and (c) improving traffic operations and efficiency, such as active traffic management. While each approach can be effective in specific contexts, their real-world applicability is often limited. Modality management strategies, although theoretically beneficial \cite{nguyen2017local, moylan2016observed}, face adoption and policy constraints that reduce their effectiveness across many major urban areas \cite{kormos2021cities}. On the other hand, capacity expansion is costly, time-consuming, and can paradoxically worsen congestion due to phenomena, such as Braess’s Paradox \cite{braess1968paradoxon}. Therefore, a large body of existing literature on optimizing network-wide vehicular traffic focuses on developing active traffic management strategies. 

Existing work on improving traffic operations and efficiency can be broadly classified into three strategies: (i) marginal cost pricing, (ii) information revelation, and (iii) vehicle routing. Marginal cost pricing, wherein drivers are charged a fee for utilization of roads in proportion to the congestion of those roads, has been shown to be theoretically and empirically effective at mitigating total network congestion \cite{vickery1969congestion, ait2021quantifying, gonzalez2023reducing}. However, these approaches have repeatedly been rejected in democratic societies due to drivers' perceptions of them being unfair or unnecessary \cite{harrington2001overcoming, gu2018congestion}, making them impractical solutions in many urban traffic networks. Similarly, information revelation strategies aiming to provide drivers with information about network conditions, have been shown theoretically to reduce traffic congestion under conditions of user compliance and homogeneous driver objectives \cite{das2017reducing, tavafoghi2017informational}. However, these  conditions rarely hold in real-world settings because drivers often shift their routes to negate predicted system-level improvements; information provided by centralized authorities is frequently delayed or incomplete \cite{arnott1991does, toso2023impact}; and many drivers disregard available information due to habitual behaviors, cognitive biases, or preferences that deviate from network-optimal routing \cite{liu2022analysis, wang2020empirical}.

Given the shortcomings of marginal cost pricing and information revelation strategies in practice, vehicle routing-based strategies have remained a central focus of research on improving traffic operations and efficiency. The aim is to influence or determine vehicle trajectories to enhance overall network performance \cite{isa2014review}. By shaping route choices, routing-based approaches directly impact how demand is distributed across the network, thereby affecting the formation and propagation of congestion. However, many classical routing strategies face practical limitations. Some raise safety concerns (e.g., those involving lane reversals \cite{li2023dynamic} or part-time shoulder use \cite{ho2023adaptive}), while others inadvertently worsen congestion due to counterintuitive effects like Braess's Paradox. Additionally, many such strategies rely on overly simplified assumptions, such as all vehicles being human-driven, fully autonomous, or part of a fixed mixture of the two, with an assumption that a subset of travelers is compliant with system routing efforts.

Since drivers are known to behave selfishly in most cases, there exists a fundamental misalignment in motives between drivers and overseeing entities that aim to manipulate traffic flows in a way that minimizes total network congestion. As such, Stackelberg game models, which capture the strategic interaction between a leader that commits to a policy, and followers that respond selfishly, offer a natural framework for modeling centralized traffic interventions in the presence of misaligned motives. In this context, Stackelberg game models allow a central planner to anticipate and influence the behavior of self-interested drivers by optimizing a routing strategy for a subset of vehicles, while accounting for the rational responses of the remaining population. These models have therefore been applied to routing interactions, resulting in Stackelberg routing games \cite{korilis1997achieving}, which explicitly model the inherent leader-follower dynamic between a central planner and self-interested agents. Indeed, traditional Stackelberg routing approaches have been shown theoretically to reduce traffic congestion in cases of static driver compliance \cite{roughgarden2001stackelberg, bonifaci2010stackelberg} as well as probabilistic driver compliance \cite{brown2024tasr}, \cite{brown2025iterative}.

Stackelberg-based approaches are well-suited for modeling interactions between selfish agents and a system-level coordinator tasked with minimizing overall congestion. However, they fall short when applied to modern traffic networks by typically assuming fixed compliance levels, static operational modes, and complete knowledge of vehicle behavior or trustworthiness \cite{roughgarden2001stackelberg, brown2024tasr, bonifaci2010stackelberg, krichene2016social, kolarich2022stackelberg}. These assumptions break down in realistic settings, including those where vehicles may switch between modalities (e.g., human control, driver assistance, tele-operation, full autonomy) during a routing interaction; the planner lacks direct access to internal beliefs and decision-making criteria of each operational modality; and the vehicle objectives may vary significantly across modalities.

While building on the potential of Stackelberg routing for congestion reduction in intelligent transportation networks, this paper extends it significantly 
to realistic traffic networks with multi-modal vehicles, where each modality optimizes individual objectives rather than system-wide metrics. To this end, this paper models the interaction between a vehicle with multiple operational modalities and a control arbitration system as a Stackelberg game and proposes a traded control allocation approach that dynamically assigns control among self-interested operational modalities to mitigate congestion.

\vspace{-0.05in}
\section{Problem Formulation \label{Sec: Problem Formulation}}
\vspace{-0.05in}

Consider a directed traffic network $\mathcal{G}=(V,E)$, where $V$ and $E$ represent the set of physical locations/intersections and the set of road segments, respectively. Each edge $e \in E$ is associated with a travel time function, $t_e(f_e)$, which depends on the flow $f_e$ traversing $e$ and captures congestion effects. As widely used, assume the road segment travel time is modeled using the bureau of public roads (BPR) function \cite{manual1964bureau}: 

\vspace{-0.1in}
\begin{equation}
t_e(f_e) = t^{ff}_e \left (1 + \lambda \left( \frac{f_e}{c_e}\right)^\beta \right),
\label{Eqn: BPR}
\end{equation}
\vspace{-0.15in}

\noindent where $t^{ff}_e$ denotes the free-flow travel time along the edge $e$, and  $c_e$ denotes the capacity of an edge. Here $\lambda$ and $\beta$ are shape coefficients, commonly assumed to be 0.15 and 4, respectively. 

Consider a scenario in which a ground vehicle with a human occupant must traverse a traffic network from an origin $v_o$ to a destination $v_d$ via a simple path, where the set of all simple paths through the network is denoted by $\mathcal{P}$ and the vehicle's induced flow on each edge in the network is given by $f_c$. Given that consumer ground vehicles are increasingly equipped with redundant perception and localization sensors to allow for multiple modes of operation (e.g., human driving,  driver assistance, remote tele-operation, full autonomy), assume the vehicle can be controlled throughout the path traversal process by one of multiple alternative operational modalities that can make routing decisions throughout the network traversal process. Let $Y = \{y_1, \cdots, y_M\}$ denote the set of $M$ operational modalities.

An vehicle control arbitration system (in short, the system), equipped with its own sensing infrastructure to perceive the network, is assumed to arbitrate among the operational modalities. The routing interaction is indexed by decision steps $k = 1, 2, \cdots, K$, where the vehicle is located at a vertex $v_k \in V$ at step $k$ and a decision must be taken by the vehicle about which edge to traverse next. 
At step $k$, the available action set is comprised of the set of outgoing edges from $v_k$, given by $E_k \subseteq E$, and an action is the selection of a single outgoing edge $e_k \in E_k$ to be traversed during step $k$.

Let $P_{k-1} =(e_1,\dots,e_{k-1})$ denote the sequence of edges comprising the path up to, but not including, step $k$. Assume that along the edges in the network, roadside units (RSUs) are present that use sensing infrastructure to observe edge flows and communicate this information to the system. At step $k$, for each candidate action $e \in E_k$, the arbitration system's uncertainty about the existing flow on $e$ is represented by a belief $\phi_{s}(\cdot \mid e, P_{k-1})$ over the nonnegative random flow $f_e$. Given that the system's beliefs are formed using information shared by the RSUs, assume that the beliefs at the system at step $k$ denote the true state of the network. Let $\Phi_s$ denote the vector of arbitration system beliefs for all edges $e \in E$. Assume that, like the system, each operational modality maintains its own prior belief about the distribution of flows on the network edges formed through information gathered by its infrastructure, denoted by $\phi_{y_m}(\cdot \mid e, P_{k-1})$ for operational modality $y_m$ at step $k$. Given that these beliefs may be derived from heterogeneous sensing infrastructures, discrepancies may arise among the operational modalities in their expected travel times and, therefore, in the edges they prefer to select. 

Following \eqref{Eqn: BPR}, the expected edge travel time induced by an action $e_k$ at step $k$ with respect to the system's belief is:

\vspace{-0.2in}
\begin{equation} 
\label{Eqn: ExpChosenEdgeTime}
\bar{t}_{k}(e_k; \Phi_s)  = \displaystyle \sum_{f_{e_k}= 0}^{\infty} t^{ff}_e \Big(1+\lambda (\tfrac{f_{e_k} + f_c}{c_e})^{\beta}\Big) \cdot \phi_{s}(f_{e_k}| e_k, P_{k-1}),
\vspace{-0.05in}
\end{equation}

\noindent while the expected travel time of an edge $e \in E$ where $e \not = e_k$ is given by

\vspace{-0.2in}
\begin{equation}
\label{Eqn: ExpEdgeTime}
\bar{t}_{k}(e; \Phi_s)  = \displaystyle \sum_{f_e = 0}^{\infty} t^{ff}_e \Big(1+\lambda (\tfrac{f_e}{c_e})^{\beta}\Big) \cdot \phi_{s}(f_e | e, P_{k-1}).
\end{equation}
\vspace{-0.1in}

\noindent The system's objective is to minimize the total expected travel time of the network. From \eqref{Eqn: ExpChosenEdgeTime} and \eqref{Eqn: ExpEdgeTime}, the instantaneous network travel time at step $k$ assuming the vehicle traverses a path $P = {e_1, \cdots, e_K}$ is given by
\begin{equation}
\label{Eqn: InstTNTT}
    T_{s,k}(P) = \bar{t}_{k}(e_k; \Phi_s) + \displaystyle \sum_{e \in E \backslash \{e_k\}} \bar{t}_{l}(e ; \Phi_s).
\end{equation}
Then the total expected network travel time is given by

\vspace{-0.2in}
\begin{equation*}
T_s(P) = \displaystyle \sum_{k = 1}^K \left[\bar{t}_{k}(e_k; \Phi_s) + \displaystyle \sum_{e \in E \backslash \{e_k\}} \bar{t}_{l}(e ; \Phi_s) \right].
\end{equation*}
On the other hand, the objective of each operational modality $y_m \in Y$ is to minimize the expected individual vehicle travel time with respect to $\Phi_{y_m}$ when traversing a path $P$. It is:

\vspace{-0.1in}
\begin{equation*}
\label{Eqn: IndVehTravelTime}
T_w(P) = \displaystyle \sum_{k=1}^K \bar{t}_k(e_k; \Phi_{y_m}).
\end{equation*}

While the system and the vehicle operational modalities have misaligned motives, the system does not have direct control over the operational modality. Instead, it is only able to allocate the vehicle's control to one of the operational modalities that may have motives and beliefs misaligned with those of the system. The interaction between the system (leader) and the set of vehicle operational modalities (followers) is modeled as a repeated Stackelberg game in which the agents possess asymmetric information about the network state. 

At each interaction step, the system allocates vehicle control to a single operational modality to determine the vehicle’s next edge decision. The system’s strategy is to alternate control among the operational modalities in such a way that the total network congestion is minimized, while anticipating that each appointed operational modality will act to minimize the vehicle's individual travel time given its own beliefs. Let $\sigma_k = \{\sigma_{1,k}, \cdots, \sigma_{M,k}\}$ denote the system's mixed strategy at each step, where $\sigma_{m,k}$ refers to the probability of allocating vehicle control to operational modality $y_m$. For ease of notation, $\sum_{m=1}^M \sigma_{m,k} = 1$, and $\sigma_{m,k} \geq 0$ for all $m \in \{1, \dots, M\}$. 

\subsection{Stackelberg Equilibrium}
In this formulation, the Stackelberg structure arises from the system committing to a distribution $\sigma_k$ over operational modalities at each step $k$. Each operational modality, when appointed, selects the outgoing edge that minimizes the vehicle’s expected individual travel time under its own beliefs $\phi_{y_m}(\cdot|e, P_{k-1})$. Given that the system does not have direct control over the path taken by the vehicle and can only allocate control to an operational modality $y_m \in Y$ at each step, the system’s objective is to choose a sequence of mixed strategies $\boldsymbol{\sigma}^* = \{\sigma_1^*, \dots, \sigma_K^*\}$ corresponding to realized active operational modalities at each step $k \in K$. The goal is to minimize the total expected travel time of the network, where $\sigma_k$ is a probability distribution over the operational modalities. 

Formally, the system seeks a strategy $\boldsymbol{\sigma}$ that minimizes the induced total expected network travel time:

\vspace{-0.15in}
\begin{equation*}
\label{Eqn: Travel Time Expectation}
    \mathbb{E}_{\phi_s}[T_{s,k}(P), \boldsymbol{\sigma}] = \displaystyle \sum_{k=1}^K \sigma_{k} \cdot T_{s,k}(P|\sigma_k),
\end{equation*}
\vspace{-0.15in}

\noindent where $P|\sigma_k$ is comprised of edges $e_{m,k}$ selected by the realized operational modality $y_m^k \sim \sigma_k$ at each step $k \in K$. The optimal system strategy is therefore given by

\vspace{-0.15in}
\begin{equation*}
\label{Eqn: SystemObjective}
    \boldsymbol{\sigma}^* = \argmin_{\boldsymbol{\sigma}} 
    \mathbb{E}_{\phi_s}[T_{s,k}(P), \boldsymbol{\sigma}].
\end{equation*}
\vspace{-0.15in}

Since the objective of each operational modality is to minimize the expected travel time of the vehicle individually, the active operational modality $y_m^k$ is assumed to choose the next edge $e^*_{k}$ by first determining the path $P_{k\to}^*(y_m^k)$ from the current vertex $v_{k}$ to the destination $v_d$. This is given by:

\vspace{-0.15in}
\begin{equation*} 
\label{Eqn: OpOptPath}
P_{k\to}^*(y_m^k) = \argmin_{P_{k\to} \in \mathcal{P}_k} \sum_{k = l}^K \bar{t}_l(e_l; \Phi_{y_m}),
\end{equation*}
\vspace{-0.1in}

\noindent where $\mathcal{P}_k$ denotes the set of all paths from vertex $v_{k}$ to $v_d$. The realized operational modality is assumed to traverse path $P_{k\to}^*$ by making the edge decision $e_{k}^* = P_{k\to}^*[1]$ at step $k$.

Given the strategies of the system and the operational modality at each step, the Stackelberg equilibrium of the game between the system and the $M$ operational modalities is characterized as follows:

\vspace{-0.25in}
\begin{equation}
\begin{array}{lll}
\label{Eqn: Game Equilibrium}
    \boldsymbol{\sigma}^* & = \displaystyle \argmin_{\boldsymbol{\sigma}} \sum_{k=1}^K T_s(P_{\boldsymbol{\sigma}}),
    \\[2ex]
    \boldsymbol{A}^*(\boldsymbol{\sigma}^*) & = \{e^*_k | y_m^k \sim \sigma_k, \; k=1,\dots,K \},
\end{array}\tag{P1}
\end{equation}
where $(\boldsymbol{\sigma}^*, \boldsymbol{A}^*(\boldsymbol{\sigma}^*))$ are the equilibrium strategies of the system and the appointed operational modalities.

The computation of Problem \ref{Eqn: Game Equilibrium} requires solving for a mixed strategy Stackelberg equilibrium at each interaction step. Given that the beliefs of the operational modalities (the followers) are unknown to the system, the routing interaction presented in this work constitutes a Bayesian Stackelberg game. Computing the Stackelberg strategy for a two-player Bayesian game is known to be NP-hard \cite{letchford2009learning}. Since this game is comprised of multiple followers, each of unknown type, in a sequential game setting in which the strategy of the leader consists of choosing a sequence of followers throughout the interaction, finding a Stackelberg strategy $\boldsymbol{\sigma}^*$ that optimizes the total network travel time is at least NP-hard. 

\vspace{-0.05in}
\section{Proposed Methodology \label{Sec: Proposed Methodology}}
\vspace{-0.05in}
Since finding an exact solution for the system's sequence of optimal strategies defined in Problem \ref{Eqn: Game Equilibrium} is NP-hard, a suboptimal 
solution is more feasible to compute in a real-world traffic network. To establish a performance baseline against which to compare TACTS, the system-optimal control allocation strategy is computed assuming the system has full knowledge of the beliefs of all operational modalities throughout the routing interaction. The total network travel time resulting from this traded control strategy is used as lower bound on the achievable total network travel time from TACTS. 

Assume the system has full knowledge of every operational modality's beliefs and can therefore perfectly predict the action each modality will take when assigned control. For every step $k \in K$, the minimum total network travel time achievable through traded control allocation among the operational modalities 
is given by
\begin{equation}
\label{Eqn: OptTNTT}
    \tau^\circledast = \! \! \min_{(m_1, \cdots, m_K) \in \mathcal{M}^K} \displaystyle \sum_{k = 1}^K \! \! \left[\bar{t}_k(e_{m,k}; \Phi_s)  + \! \! \! \! \displaystyle \sum_{e \in E \backslash\{e_{m,k}\}} \! \! \! \! \bar{t}_k(e; \Phi_s) \! \right],
\end{equation}
where $m_{1:K}=(m_1,\dots,m_K)\in\mathcal{M}^K$ denotes a sequence of deterministic modality assignments over the step horizon $K$, and $e_{m,k}$ denotes the edge decision made by the realized operational modality $y_m^k$ at step $k$.

While $\tau^\circledast$ serves as a performance baseline for the strategy $\boldsymbol{\sigma}^s$ chosen by the system, in reality, the system would need to learn the beliefs of each operational modality throughout repeated routing interactions. As such, the proposed solution approach utilizes a trust-aware, regret matching-based traded control algorithm to determine the system's traded control decision among the set of vehicle operational modalities at each step. In the developed approach, operational modalities whose decisions deviate from the system’s anticipated best actions are penalized, while those that align more closely are favored. This approach yields a sequential update of the system's mixed strategy based on historic operational modality performance. Additionally, employing a mixed strategy rather than a pure strategy allows for exploration of strategies at each step.

\vspace{-0.05in}
\subsection{Trustworthiness Model \label{Sec: Trust Model}}
\vspace{-0.05in}
Trust has long been interpreted as a dynamic belief about the historic reliability of an agent \cite{gambetta2000can, wang2018automation}.  Following this perspective, the system is assumed to compute the trustworthiness of each operational modality by aggregating a regret score for having relied on that operational modality throughout the routing interaction. Conceptually, the trustworthiness of an operational modality corresponds to the probability of it having a belief aligned with that of the system. Let $\alpha_{m,k-1} \in \mathbb{R}$ denote the system’s estimated relative trustworthiness score for modality $y_m^{k}$, where relative trustworthiness is considered rather than absolute trustworthiness to guarantee that the vehicle is always controlled by one of the operational modalities. 

\begin{figure}[htbp]
\vspace{-0.05in}
\centering
\includegraphics[width=0.48\textwidth]{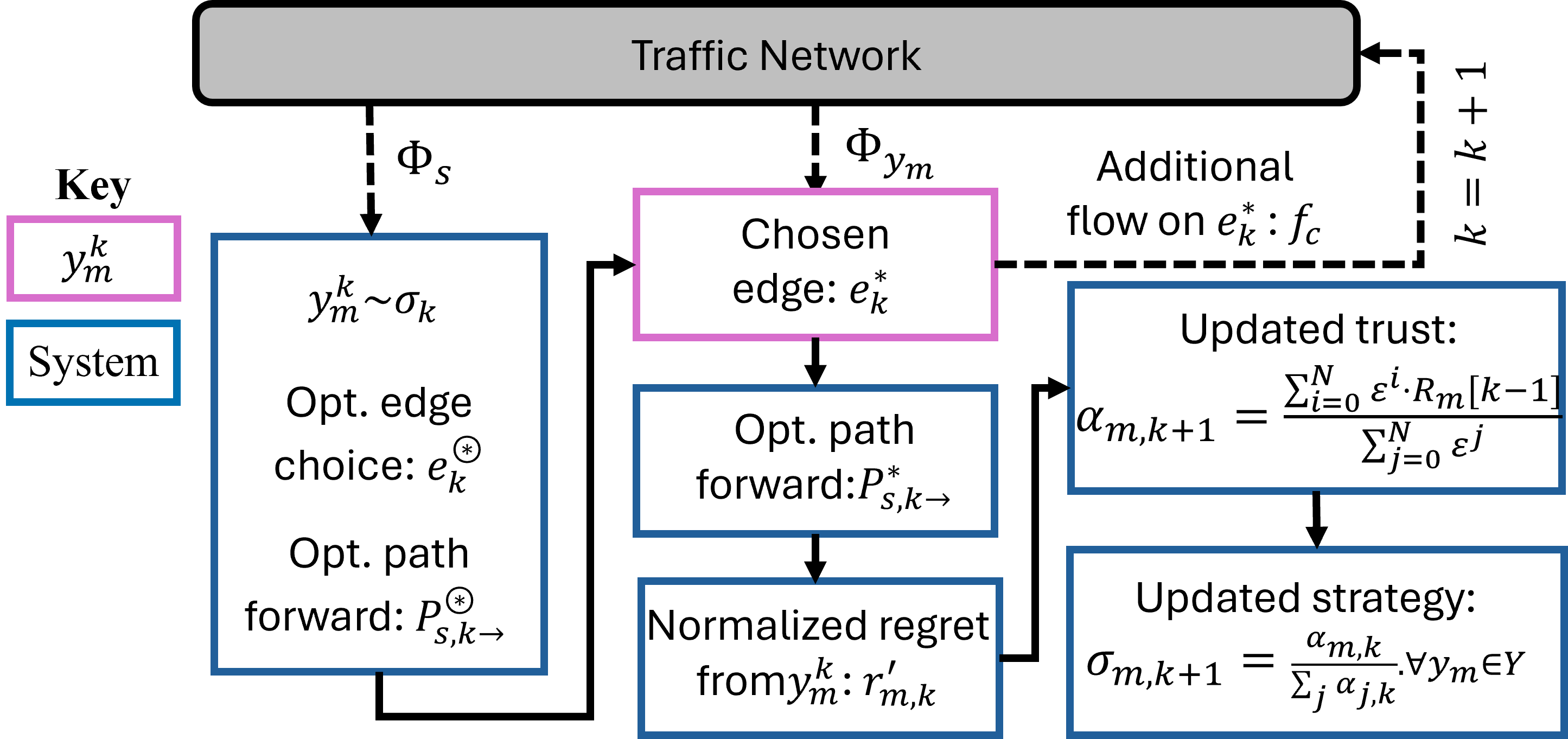}
\caption{Overview of key steps of the TACTS algorithm.}
\label{Img: Algorithm Overview}
\vspace{-0.2in}
\end{figure}

Let $P^*_{s, k\to}$ denote the system-optimal path through the remainder of the network given that edge $e_{k}^*$ was traversed at step $k$, given by
\begin{equation*}
T_s(P_{s,k\to}^*) =
 \displaystyle \sum_{l = k}^K \left[\bar{t}_{s,l}(e_l^*; \Phi_s) + \! \! \! \! \! \displaystyle \sum_{e \in E \backslash \{e_l^*\}} \! \! \! \! \! \bar{t}_{s,l}(e ; \Phi_s) \right].
\end{equation*}
Let $e_{k}^\circledast$ denote the edge that would have been traversed at step $k$ if the system had direct control over the edge decisions taken by the vehicle at the previous step, which is given by $T_s(P_{s,k-1\to}^*)[1]$. Let $P_{s, k\to}^\circledast$ denote the system-optimal path through the remainder of the network assuming that edge $e_{k}^\circledast$ had been taken at step $k$. The system's instantaneous regret for relying on operational modality $y_m^k$ is then given by 
\begin{equation}
\label{Eqn: System Instantaneous Regret}
r_{m,k} = T(P^*_{s, k\to}) - T(P_{s, k\to}^\circledast).
\end{equation}

Positive regret indicates that the operational modality’s action incurred a greater expected travel time relative to the system’s preferred action with respect to the system's belief. Note that, from  \eqref{Eqn: System Instantaneous Regret}, the minimum instantaneous regret is $r_{m,k} = 0.0$, which occurs when $e_{k}^* = e_{k}^\circledast$. On the other hand, the maximum instantaneous regret is given by 
\begin{equation*}
    r_{m,k}^\circleddash = T_s(P_{s,k\to}^\circleddash) - T(P_{s, k\to}^\circledast).
\end{equation*}
Here, $T_s(P_{s,k\to}^\circleddash)$ is the total network travel time resulting from an edge decision $e_k^\circleddash = P_{s,k\to}^\circleddash[1]$ that would maximize congestion for the remainder of the routing interaction, where 
\begin{equation*}
P_{s,k\to}^\circleddash = \argmax_{P_k \in \mathcal{P}_k} \displaystyle \sum_{l = k}^K \left[\bar{t}_{s,l}(e_l^\circleddash; \Phi_s) + \! \! \! \! \! \displaystyle \sum_{e \in E \backslash \{e_l\}} \! \! \! \! \! \bar{t}_{s,l}(e ; \Phi_s) \right].
\end{equation*}
To ensure that the updated trustworthiness score of operational modality $y_{m}^k$ is in $[0,1]$, let the normalized instantaneous regret be computed as

\vspace{-0.2in}
\begin{equation*}
r_{m,k}^\prime = \frac{r_{m,k}}{r_{m,k}^\circleddash}.
\end{equation*}

\vspace{-0.05in}
Let $R_m$ denote the vector of normalized regret values $r_{m,k}^\prime$ of modality $y_m^k$. Following the computational trust model of \cite{wang2022computational}, the system’s trust in operational modality $y_m^k$ is computed as an exponentially weighted average of past normalized regrets, given by
\begin{equation}
\label{Eqn: Trust Update}
\alpha_{m, k+1} = \frac{\sum_{i=0}^{N} \varepsilon^i \cdot (1 - R_{m}[k-i])}{\sum_{j=0}^N \varepsilon^j},
\end{equation}
where $\varepsilon \in (0,1)$ controls memory decay, and $N \leq k$ denotes the horizon of most recent steps considered. In this formulation, lower regret corresponds to higher trust, meaning that operational modalities consistently aligned with the system’s best actions will have higher trustworthiness scores.

Unlike conventional regret-matching approaches in which the strategies (or trustworthiness scores) of all agents are updated at each iteration using counterfactual information, the proposed algorithm updates the trustworthiness score of only the active operational modality at the current step. This design reflects the realistic constraint that the system does not have access to the counterfactual decisions of modalities that are not active or have not been active and cannot make predictions about their unobserved behavior. As a result, the approach is a regret matching-based algorithm that operates under partial information, where trust updates are based exclusively on observed actions.

\vspace{-0.05in}
\subsection{Traded Control Methodology \label{Sec: Defense Methodology}}
To determine the active operational modality at each step, the system employs a regret matching-based algorithm \cite{hart2000simple}, augmented with trustworthiness estimates. Let $\sigma_0$ be the initial uniform distribution over operational modalities, with neutrality assumed at the outset of the routing interaction:

\vspace{-0.1in}
\begin{equation*}
\sigma_0(y_m) = \frac{1}{M}, \quad \forall \ y_m \in Y,
\end{equation*}
and assume initial trustworthiness scores are given by $\alpha_{m,0} = \sigma_0(y_m)$.  

At each step $k$, the system samples an operational modality $y_m^k \sim \sigma_k$, observes the edge choice $e_k^*$, determines an optimal path $P_{s,k\to}^*$ from the $v_k$ to $v_d$, and computes a system-optimal edge choice $e_k^\circledast$ from the system-optimal path forward, $P_{s,k\to}^\circledast$. By comparing the total network travel times of $P_{s,k\to}^*$ and $P_{s,k\to}^\circledast$, the system then computes a normalized regret score $r_{m,k}^\prime$ via \eqref{Eqn: System Instantaneous Regret}, and updates the trustworthiness score $\alpha_{m,k+1}$ for the active operational modality using \eqref{Eqn: Trust Update}. Finally, the system updates its mixed strategy $\sigma_{k+1}$ by reweighting probabilities according to the updated trustworthiness scores as follows:

\vspace{-0.1in}
\begin{equation}
\label{Eqn: Mixed Strategy Update}
\sigma_{k+1}(y_m) = 
    \displaystyle \frac{\alpha_{m,k+1}}{\sum_{j} \alpha_{j,k+1}}, \quad \forall y_m \in Y.
\end{equation}
\vspace{-0.1in}

To ensure that the trustworthiness score of each operational modality remains normalized and relative to all other operational modalities, assume at the conclusion of step $k$, $\alpha_{m,k+1} = \sigma_{k+1}(y_m)$ for all $y_m \in Y$. The proposed trust update and traded control method result in the system gradually shifting its mixed strategy towards favoring more trustworthy operational modalities, 
with the inertia of the system's mixed strategy being controlled by the trustworthiness of each operational modality. The developed algorithm, 
TACTS, is shown in Algorithm \ref{Alg: Traded Control Alg}, and a high-level overview of key algorithmic steps is provided in Fig. \ref{Img: Algorithm Overview}. An example of the proposed algorithm in practice is provided in Section \ref{Sec: Example}.

\vspace{-0.05in}
\subsection{An Illustrative Example} \label{Sec: Example}
\vspace{-0.05in}
Suppose a simple traffic network is comprised of six edges: $(a,b)$, $(b,c)$, $(c,d)$, $(a,e)$, $(e,f)$ and $(f,d)$, with two distinct paths, $P_1 = [(a,b),(b,c),(c,d)]$ and $P_2 = [(a,e), (e,f),(f,d)]$. Assume a vehicle seeks a path from $v_o = a$ to $v_d = d$, and the vehicle has two operational modalities. For the sake of example, assume $N = 1$ and $\varepsilon = 0.1$. At the outset of the routing interaction ($k = 0$), $\sigma_0 = \{0.5, 0.5\}$. Suppose the system samples $\sigma_0$, resulting in the second modality being $y_m^0$, and assume $y_m^0$ has a belief vector $\Phi_{y_m}$ such that the anticipated path travel time of the vehicle is 30 minutes along $P_1$ and 24 minutes along $P_2$, making $P_2$ the selfish path of $y_m^0$, with an edge decision of $e_0^* = (a,b)$. On the other hand, assume the system has a belief vector $\Phi_s$ such that the expected total network travel time of $P_1$ is 78 minutes, compared to 83 minutes for $P_2$. Given that $P_1$ is the system-optimal path forward (i.e., $P_{s,0\to}^\circledast$), $e_0^\circledast = (a,e)$, and the instantaneous regret for allocating control to $y_m^0$ is $r_{m,0} = 83 - 78 = 5$. Note that $P_2$ is the path forward through the network leading to the worst-case total network travel time, so the normalized regret is given by $r_{m,0}^\prime = \frac{5}{5} = 1$. Since $R_m[0] = 1$, $N = 1$, and $|R_m| = 1$, the updated trustworthiness score $\alpha_{m,1} = \frac{0.1^0 \cdot (1 - 1)}{0.1^0} = \frac{0}{1} = 0,$ and the system's updated mixed strategy is $\sigma_1 = \{\frac{0.5}{0.5}, \frac{0}{0.5}\} = \{1, 0\}$. This process is repeated for the remaining steps, incrementing $k$ at each step, until $v_d$ is reached.

\begin{algorithm}[!t]
\caption{TACTS: Trust-Aware Control Trading Strategy}\label{Alg: Traded Control Alg}
\begin{algorithmic}[1]
\State $\sigma_0(y_m) = \tfrac{1}{M}$, \ $\forall y_m \in Y$ 
\State $\alpha_{m,0} = \sigma_0(y_m)$, \ $\forall y_m \in Y$ 
\For{$k = 1 \ \textbf{to} \ K$}
    \Statex \textit{// Update system belief based on observed network state}
    \State{$\Phi_{s} \gets$ \Call{UpdateBelief}{$\Phi_{s}$, $P_{k-1}$}}
    
    \Statex \textit{// Compute system-optimal path and action from $v_k$}
    \State{$P_{s,k\to}^\circledast \gets$ \Call{GetBestPath}{$\Phi_{s}, P_{k-1}$}}
    \State{$e_k^\circledast \gets P_{s,k\to}^*[1]$}

    \Statex \textit{// Sample modality according to mixed strategy}
    \State{$y_m^k \sim \sigma_k$}
    \Statex \textit{// Observe operational modality action}
    \State{$e_{k}^* \gets$ \Call{ModalityAction}{$y_m^k, P_{k-1}$}}

    \Statex \textit{// Compute projected path forward}
    \State{$P_{s,{k}\to}^* \gets$ \Call{GetProjectedPath}{$e_{k}^*, P_{k-1}$, $\Phi_{s}$}}

    \Statex \textit{// Compute worst path forward}
    \State{$P_{s,{k}\to}^\circleddash \gets$ \Call{GetWorstPath}{$\Phi_{s}, P_{k-1}$}}
    
    \Statex \textit{// Compute system regret}
    \State{$r_{m,k} \gets$ \Call{ComputeRegret}{$P_{s,{k}\to}^*$, $P_{s,{k}\to}^\circledast$, $\Phi_{s}$}}

    \Statex \textit{// Compute normalized regret}
    \State{$r_{m,k}^\prime \gets$ \Call{NormalizeRegret}{$r_{m,k}$, $P_{s,k\to}^\circleddash$}}
    
    \Statex \textit{// Update operational modality trustworthiness}
    \State{$\alpha_{m,k+1} \gets$ \Call{UpdateTrust}{$\alpha_{m,k}$, $r_{m,k}^\prime$, $N$, $\varepsilon$}}
    
   \Statex \textit{// Update system mixed strategy}
    \State{$\sigma_{k+1} \gets$ \Call{UpdateStrategy}{$\{\alpha_{j,k}\}_{j=1}^M$, $\{\boldsymbol{R}_j\}_{j=1}^M$}}

\EndFor
\State \Return{$\{\sigma_k\}_{k=1}^K$}
\end{algorithmic}
\end{algorithm}

\vspace{-0.05in}
\section{Theoretical Analysis}
\label{Sec: Theoretical Results}
\vspace{-0.05in}
To theoretically quantify the performance of the developed TACTS algorithm, this section analyzes the realized total network travel time $\tau(\boldsymbol{\sigma}^s)$ achieved under the traded control allocation strategy computed by TACTS, compares $\tau(\boldsymbol{\sigma}^s)$ to the system-optimal total network travel time $\tau^\circledast$, 
and provides an upper bound on $\tau(\boldsymbol{\sigma}^s)$. 

\begin{lemma}
\label{Lem: TACTS}
    The total network travel time achieved by the strategy $\boldsymbol{\sigma}^s$ found using the TACTS algorithm is given by 
    \begin{equation}
    \label{Eqn: TACTSTNTT}
    \tau(\boldsymbol{\sigma}^s) = \displaystyle \sum_{k=1}^K \left[t_{e_k^*}(f_{e_k^*} + f_c) + \displaystyle \sum_{e \in E \backslash \{e_k^*\}} t_e(f_e) \right].
    \end{equation}
    \begin{proof}
        Let $\tau(\boldsymbol{\sigma}^s)$ denote the realized total network travel time resulting from a traded control allocation strategy $\boldsymbol{\sigma}^s$. Given that $e_k^*$ denotes the edge decision made by the realized operational modality $y_m^k \sim \sigma_k$, to compute the realized total network travel time resulting from $\boldsymbol{\sigma}^s$, the realized network travel time at step $k$ resulting from the existing flow $f_{e_k^*}$ on edge $e_k^*$ and the vehicle's flow $f_c$ is computed and summed with the realized travel time of the existing flow $f_e$ on all other network edges. Therefore, summing the realized network travel time at step $k$ across all steps $k = 1, \cdots, K$ yields the realized total network travel time $\tau(\boldsymbol{\sigma})^s$. Hence the proof. 
    \end{proof}
\end{lemma}

Given the result in Lemma \ref{Lem: TACTS} quantifying the performance $\tau^(\boldsymbol{\sigma}^s)$ of the traded control allocation strategy $\boldsymbol{\sigma}^s$ computed using the TACTS algorithm, the following definition compares $\tau(\boldsymbol{\sigma}^s)$ to the optimal total network travel time $\tau^\circledast$ achievable through traded control among the operational modalities. 

\begin{definition}
\label{Def: PerformanceRatio}
    The \textbf{performance ratio} of the TACTS algorithm is the ratio of the realized total network travel time resulting from the traded control allocation strategy found via the TACTS algorithm and the total expected network travel time assuming the system had complete knowledge of the beliefs of each operational modality at each time step $k = 1, \cdots, K$. In other words, following \eqref{Eqn: TACTSTNTT} and \eqref{Eqn: OptTNTT}, the performance ratio of TACTS is $\displaystyle \frac{\tau(\boldsymbol{\sigma}^s)}{\tau^\circledast}.$
\end{definition}

\begin{lemma}
    A traded control allocation strategy $\boldsymbol{\sigma}$ is optimal if it results in a realized total network travel time $\tau^*$ such that 
    \begin{equation}
        \frac{\tau(\boldsymbol{\sigma}^s)}{\tau^\circledast} = 1.
    \end{equation}
    \begin{proof}
        From \eqref{Eqn: OptTNTT}, $\tau^\circledast$ denotes the minimum total expected network travel time achievable through a traded control allocation strategy among the operational modalities throughout all steps of the routing interaction $k \in K$. Necessarily, $\tau(\boldsymbol{\sigma}^s) \geq \tau^\circledast.$ Therefore, if $\tau(\boldsymbol{\sigma}^s) = \tau^\circledast$, the strategy $\boldsymbol{\sigma}^s$ achieves this minimum, leading the ratio of $\tau(\boldsymbol{\sigma}^s)$ and $\tau^\circledast$ to be $\frac{\tau(\boldsymbol{\sigma}^s)}{\tau^\circledast} = 1$, which was to be shown.
    \end{proof}
\end{lemma}

\begin{theorem}
\label{Thm: TACTSEquilibrium}
Assuming the system's belief vector $\Phi_s$ accurately reflects the true network state, the realized total network travel time $\tau(\boldsymbol{\sigma}^s)$ satisfies
\begin{equation}
\label{Eqn:TACTSBoundCorrected}
\tau^\circledast \leq \tau(\boldsymbol{\sigma}^s) \leq \tau^\circledast + \Delta_{\mathrm{regret}},
\end{equation}
where $\tau^\circledast$ is the system-optimal total network travel time and $\Delta_{\mathrm{regret}}$ is a loose upper bound on the cumulative deviation from the system-optimal total network travel time incurred by relying on the active operational modalities, defined as the sum of normalized instantaneous regrets given by
\begin{equation*}
\Delta_{\mathrm{regret}} = \sum_{m=1}^M \sum_{k=1}^K r_{m,k}.
\end{equation*}

\begin{proof}
From \eqref{Eqn: System Instantaneous Regret}, at each step $k$, the instantaneous regret $r_{m,k}$ captures the difference in total network travel time for the remainder of the path if the system had chosen the previous edge based on $\Phi_s$ compared to the active operational modality’s decision. The total network travel time under $\tau(\boldsymbol{\sigma}^s)$ is the sum of travel times over all edges actually traversed by the vehicle. Since each $r_{m,k}$ loosely upper-bounds sub-optimality induced by the active operational modality at step $k$, including future travel time from that step onward, 
$\Delta_{\mathrm{regret}} = \sum_{m=1}^M \sum_{k=1}^K r_{m,k}$ provides a loose upper bound on the deviation of $\tau(\boldsymbol{\sigma}^s)$ from $\tau^\circledast$. Thus, the realized total network travel time satisfies Inequality \eqref{Eqn:TACTSBoundCorrected}. Hence the proof.  
\end{proof}
\end{theorem}

\vspace{-0.05in}
\section{Performance Evaluation\label{Sect: Performance Evaluation}}
\vspace{-0.05in}
\begin{figure*}[!t]
    \centering
    \vspace{-0.1in}
    \begin{subfigure}[b]{0.32\textwidth}
        \centering
        \includegraphics[width=\textwidth]{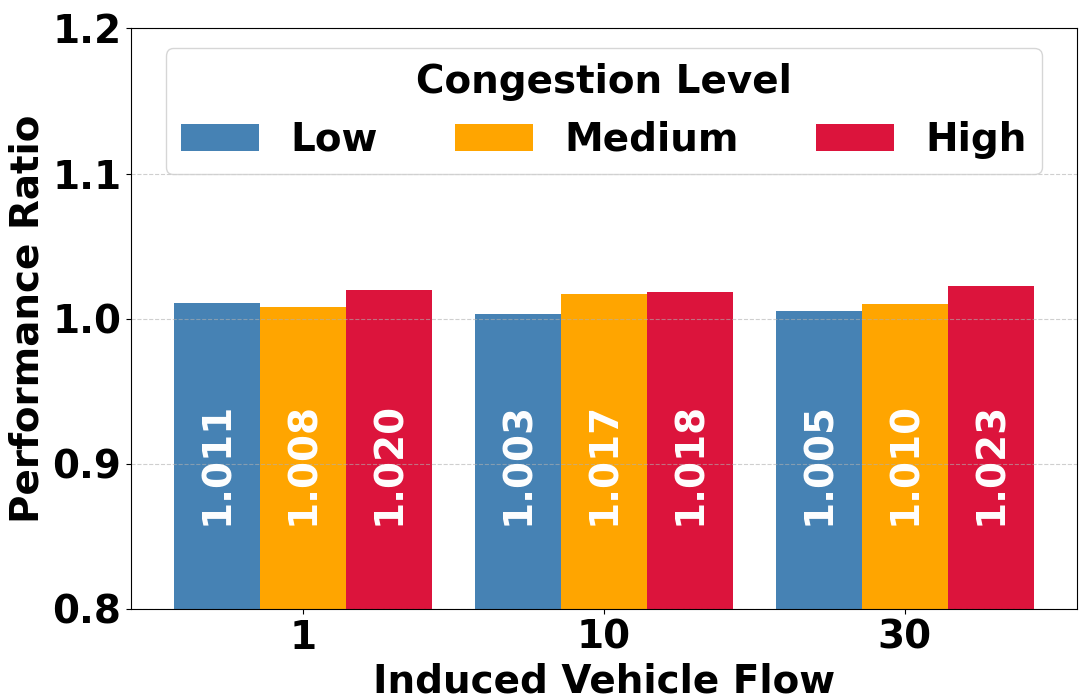}
        \caption{Anaheim (small)}
        \label{fig:anaheim}
    \end{subfigure}
    \hfill
    \begin{subfigure}[b]{0.32\textwidth}
        \centering
        \includegraphics[width=\textwidth]{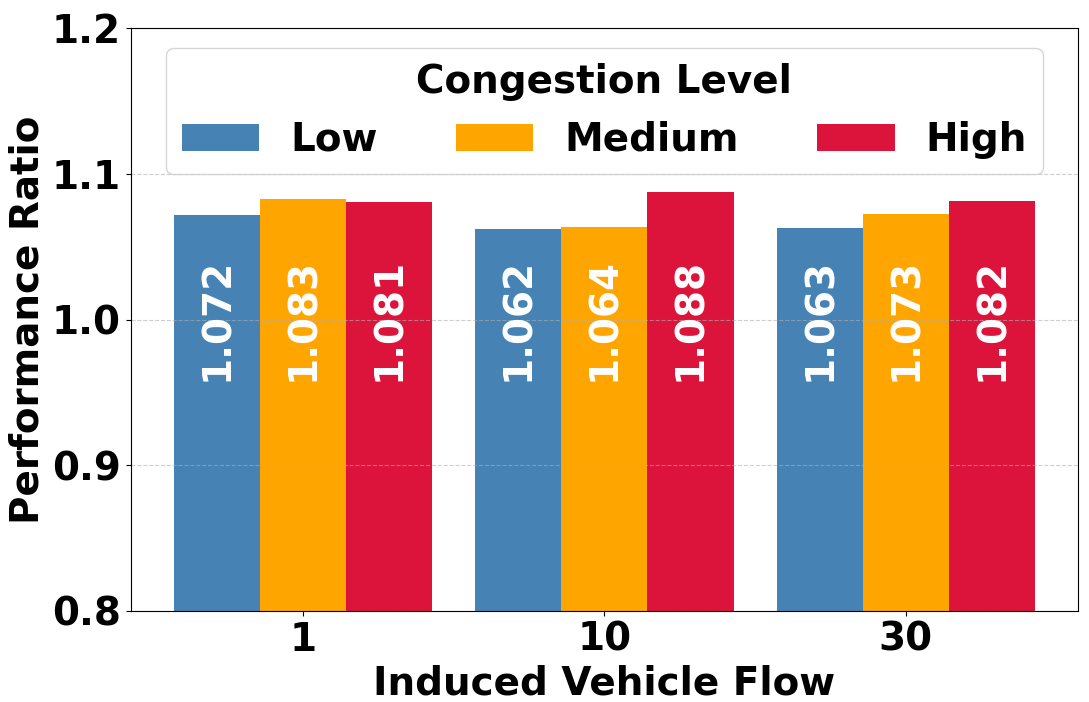}
        \caption{Sioux Falls (medium)}
        \label{fig:siouxfalls}
    \end{subfigure}
    \hfill
    \begin{subfigure}[b]{0.32\textwidth}
        \centering
        \includegraphics[width=\textwidth]{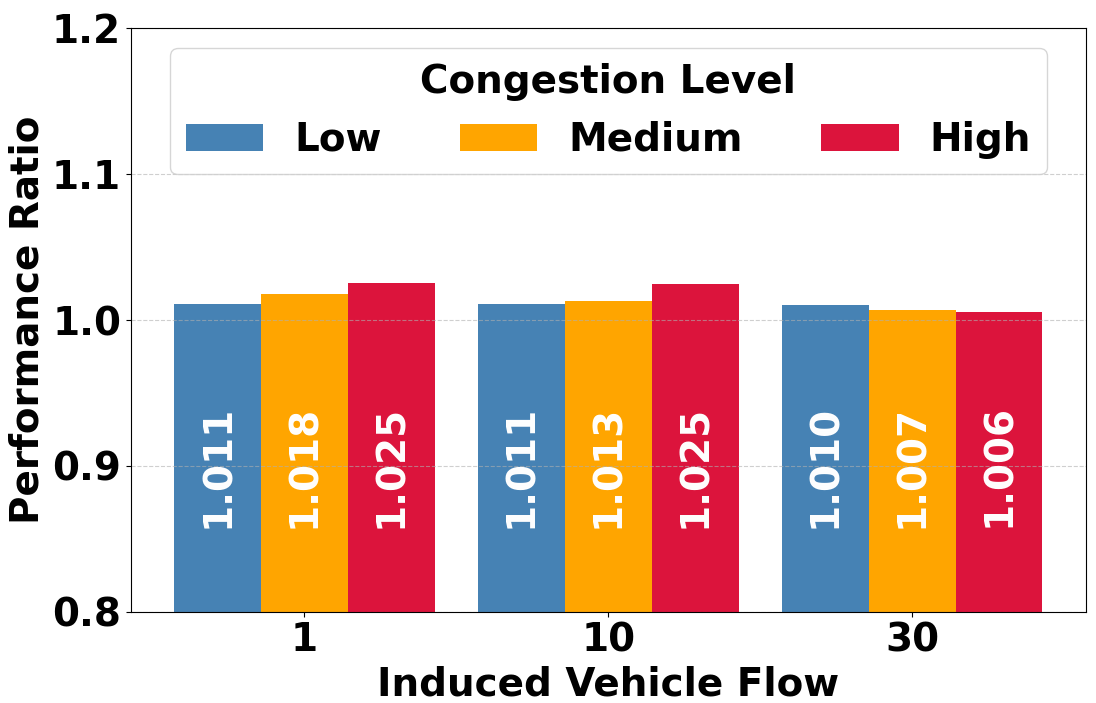}
        \caption{Chicago Sketch (large)}
        \label{fig:chicago}
    \end{subfigure}
\captionsetup{justification=centering, font=footnotesize}
    \vspace{-0.05in}
    \caption{Comparison of performance ratio of TACTS under low, medium, and high congestion levels for induced vehicle flows $f_c \in \{1.0, 10.0, 30.0\}$ across three networks: Anaheim, Sioux Falls, and Chicago Sketch.}
    \label{fig:all_networks}
    \vspace{-0.2in}
\end{figure*}

\subsection{Performance Metrics}
\vspace{-0.05in}
For each network considered, the primary performance metric was the total network travel time (i.e., congestion). 
To assess the practical applicability of the TACTS algorithm in intelligent transportation systems, the execution time was also considered as a performance metric.

To establish a performance benchmark, a degenerate case of the control-switching algorithm based on degrees of conflict (DOC) between an autonomous vehicle and a human driver, as introduced by Sourav and Cheng \cite{sourav2024work}, was adapted for comparison. The DOC algorithm originally assumes only two modalities and direct access to modalities' sensor information, whereas the scenario considered here allows for multiple operational modalities and assumes the arbitration system cannot directly observe internal vehicle information. To better fit this more realistic setting, the algorithm is assumed to default to an initial modality (analogous to an autonomous controller). The resulting total network travel time from the edge decisions of this modality over a fixed window tracked, while the system also tracks decisions it would have preferred if it had direct control of the vehicle at each step. The DOC score is given by the number of differences between the decision of the active modality and the preferred decision of the system across the time window. If this DOC score exceeds a nonzero threshold $\gamma$, control is reassigned uniformly at random to another operator. 

A degenerate version of the state-of-the-art Stackelberg routing algorithm, Trust-Aware Stackelberg Routing (TASR), was also adapted for comparison. In this simplified formulation, the system-optimal path is computed once at the outset of the routing interaction based on predicted trustworthiness levels of the available operational modalities, with no intermediate strategy updates performed during execution. Control is then assigned to the modality predicted to be most trustworthy, 
while all other modalities are assumed to behave selfishly. This formulation reflects a setting in which no direct information exchange with the vehicle’s internal decision-making processes is possible, and thus no real-time routing recommendations can be issued.

To further contextualize the performance of TACTS algorithm, two additional baseline strategies were implemented: Single Control (SC), in which a single operational modality retains control for the entire routing interaction; and Random Control Strategy (RCS), in which control is reassigned uniformly at random from all modalities at each step. The SC policy can be interpreted as allowing one modality (e.g., a human driver) to maintain control throughout the routing interaction regardless of routing decisions taken.

\vspace{-0.05in}
\subsection{Experiments, Datasets, Preprocessing} \label{Sect: Experiments, Datasets, Preprocessing}
\vspace{-0.05in}
To evaluate the effectiveness and practical feasibility of TACTS, two experiments designed to test its performance across different operating conditions were conducted. Experiment 1 assesses how TACTS scales with network size and topology and how closely it approaches the system-optimal total network travel time under different congestion levels by simulating traffic dynamics on three real-world transportation networks: Anaheim (California, USA), Sioux Falls (South Dakota, USA), and Chicago Sketch (Illinois, USA). Structurally, these correspond to small, medium, and large networks. For each network, total network travel time was evaluated under low, medium, and high congestion scenarios. 

Experiment 2 was designed to examine the adaptive behavior and comparative performance of TACTS relative to DOC, TASR, RCS, and SC under varying operating conditions. The Sioux Falls network was considered exclusively, and repeated routing interactions were simulated across low, medium, and high congestion levels with varying induced vehicle flows of $f_c \in \{1.0, 10.0, 30.0\}$. 
In addition to total network travel time, this experiment also tracked execution time to assess the practical deployability of TACTS in real-time decision-making environments.

For each network, all possible commodities comprised of nonempty path sets were considered as potential commodities to be traversed. Commodity paths were identified using a breadth-first search avoiding cycles and limiting path lengths to fewer than six edges for computational tractability. For a given routing instance, a single commodity was selected uniformly at random to represent the origin $v_o$ and destination $v_d$ of a vehicle interacting with the arbitration system. To model low, moderate, and high existing traffic congestion in the network, existing edge flows were assumed to be equal to the edge capacity scaled by 0.25, 0.75, and 1.5, respectively. Flow scaling factors were chosen based on their consequences in the BPR function, 
where edge travel times increase as the ratio of edge flow to capacity increases.

At the outset of each experiment, the system's initial trust estimate for each operational modality was assumed to be uniform, with a value of $1 / M$ for each modality. For each network edge, the true flow value was drawn from an underlying true edge flow distribution. Each operational modality was assigned a true trustworthiness score in the range $[0, 1]$. For each network edge, an operational modality's expected belief about the edge flow was generated probabilistically based on this trustworthiness score. Each operational modality was assumed to perfectly estimate true edge flow values with a probability equal to the modality's true trustworthiness. With the complementary probability, the operational modality's expected edge flow was assumed to either be lower or higher than the true flow, with a degree of uncertainty proportional to the modality's true trustworthiness. Upper and lower bounds for this amount of perturbation were computed by computing the addition (or subtraction, respectively) of true expected edge flow and a noisy version of the edge capacity multiplied by 1.5 (i.e., the edge flow under high congestion). Here, the noise scale was computed by raising the complement of the operational modality's true trustworthiness to a power of 0.5, to ensure that lower trustworthiness scores resulted in higher variance in belief. Given the lower and upper bounds computed, the modality's expectation of the edge flow was determined by randomly sampling from a uniform distribution bounded between these values. 

A history window of $N = 2$ and a memory decay parameter of $\varepsilon = 0.01$ were used for all experiments. Each experiment was repeated 500 times per network, congestion level, and induced flow setting. For each repetition, the number of operational modalities was sampled uniformly from the range $[2, 6]$ to reflect plausible real-world vehicle capabilities. Experimentation was conducted using Python 3.12.6 on a machine using an AMD Ryzen 9 7945HX processor with 32 GB of DDR5 RAM in a single-threaded environment.

\begin{figure}[htbp]
\vspace{-0.1in}
\centering
\includegraphics[width=0.48\textwidth]{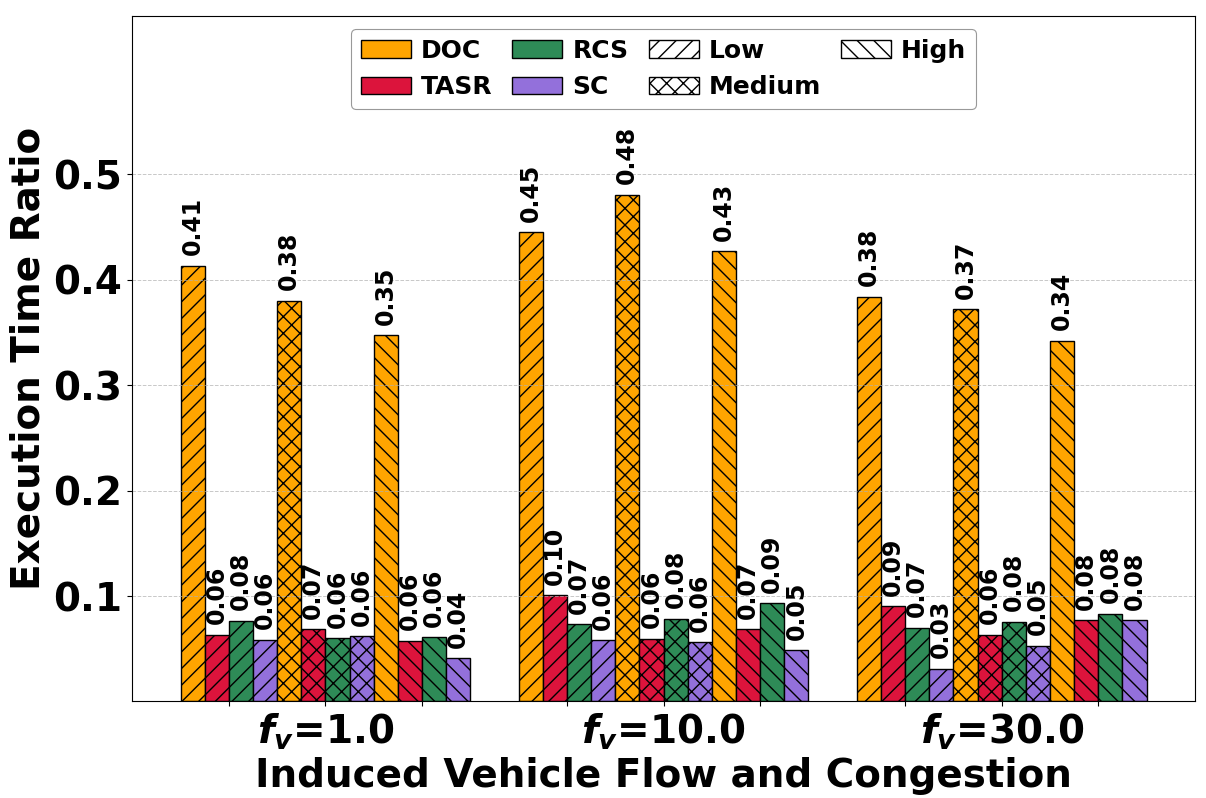}
\vspace{-0.05in}
\captionsetup{justification=centering, font=footnotesize}
\caption{Average ratio of execution time compared to TACTS in Sioux Falls network across low, medium, and high congestion levels for induced vehicle flows $f_c \in \{1.0, 10.0, 30.0\}$.}
\label{Img: Avg Execution Ratio}
\vspace{-0.25in}
\end{figure}

\vspace{-0.1in}
\subsection{Experimental Results \label{Sect: Results}}
\vspace{-0.05in}
The results of Experiment 1, shown in Fig. \ref{fig:all_networks}, report the average performance ratio (Definition \ref{Def: PerformanceRatio}) derived from the total network travel times after executing the TACTS algorithm on the Anaheim, Sioux Falls, and Chicago Sketch networks under low, medium, and high congestion for various induced vehicle flows. Across networks with substantially different sizes and topologies, TACTS consistently operated near the system-optimal total network travel time baseline, even when induced vehicle flows were high. In all scenarios, the average performance ratio was within $8.76\%$ of the optimal performance ratio, with the closest result deviating by only $0.5\%$. Thus, TACTS maintains strong scalability and robustness, delivering near-optimal performance across heterogeneous traffic environments. 

Results of Experiment 2, presented in Table \ref{tab: MC Travel Time}, further demonstrate the effectiveness of the TACTS algorithm relative to alternative control-switching and routing strategies in the Sioux Falls network across a variety of congestion levels and induced vehicle flows. TACTS consistently achieves the lowest or near-lowest performance ratio in the majority of cases, with its relative performance tending to become more pronounced as induced vehicle flows and congestion increase. On the other hand, alternative strategies, notably DOC and TASR, that tend to perform well under low induced vehicle flows show an increase in inefficiency as induced vehicle flow increases beyond atomic flows. Note that as the induced vehicle flows and congestion increase, the system-optimal total network travel time increases, resulting in an increased impact of the TACTS algorithm in comparison to each alternative strategy. For example, under high congestion when $f_c = 1.0$, DOC offers only $0.65\%$ performance improvement over TACTS, while when congestion is high under $f_c = 30.0$, TACTS offers a performance improvement of $12.35\%$ over DOC.

\begin{table}[t]
\caption{Average performance ratio for each algorithm in Sioux Falls network for induced vehicle flows $f_c \in \{1.0, 10.0, 30.0\}$ under varied congestion. Values closest to the system-optimal total network travel time are \textbf{bold}.}
\vspace{-0.05in}
\label{tab: MC Travel Time}
\centering
\begin{tabular}{c c c c c c c c}
\hline
\\[-1.5ex]
\multicolumn{1}{c}{} & \textbf{Congestion} & \textbf{TACTS} & \textbf{DOC} & \textbf{TASR} & \textbf{RCS} & \textbf{SC}
\\[0.5ex]
\hline\hline
\\[-2ex]
 & $Low$ & \textbf{1.072} & 1.078  & 1.084 & 1.073 & 1.068
\\[0.5ex]
\multirow{2}*{\textbf{$f_c = 1$}} & $Medium$ & 1.083 & \textbf{1.069} & 1.081 & 1.069 & 1.066
\\[0.5ex]
& $High$ & 1.081 & \textbf{1.074} & 1.079 & 1.080 & 1.090
\\[0.5ex]
\hline
\\[-2ex]
 & $Low$ & \textbf{1.062} & 1.070 & 1.076 & 1.072 & 1.073
\\[0.5ex]
\multirow{2}*{\textbf{$f_c = 10$}} & $Medium$ & \textbf{1.064} & 1.076 & 1.076 & 1.075 & 1.069
\\[0.5ex]
& $High$ & \textbf{1.088} & 1.110 & 1.090 & 1.090 & 1.090
\\[0.5ex]
\hline
\\[-2ex]
 & $Low$ & \textbf{1.063} & 1.078 & 1.069 & 1.069 & 1.065
\\[0.5ex]
\multirow{2}*{\textbf{$f_c = 30$}} & $Medium$ & \textbf{1.073} & 1.074 & 1.077 & 1.078 & 1.075
\\[0.5ex]
& $High$ & \textbf{1.081} & 1.091 & 1.082 & 1.098 & 1.089
\\[0.5ex]
\hline
\end{tabular}
\vspace{-0.13in}
\end{table}

From Experiment 2, the average ratio of the execution time of each alternative strategy is compared to that of TACTS in Fig. \ref{Img: Avg Execution Ratio}, showing that the observed performance gains come at the cost of moderately higher computational overhead for TACTS. However, this trade-off may be considered negligible in practice for real-time decision-making in intelligent transportation systems, since execution times are on the order of milliseconds. As such, the improved routing efficiency and network-level performance achieved by TACTS, compared to alternative strategies, are realized without sacrificing operational responsiveness or deployability.

\vspace{-0.1in}
\section{Conclusion
\label{Sect: Conclusion}}
\vspace{-0.0in}
This paper represents the increasing prevalence of ground vehicles capable of dynamically switching between multiple operational modalities rather than being exclusively human-driven or fully autonomous. Specifically, it models the interaction between a vehicle control arbitration system and a vehicle with multiple alternative self-interested operational modalities as a Bayesian Stackelberg game. In his game model, the system aims to construct a traded control strategy that mitigates total network travel time in a transportation network, despite misaligned motives with those of the operational modalities. 

Building on this formulation, a trust-aware control trading strategy based on regret-minimization, referred to as the TACTS algorithm, is developed. This strategy dynamically allocates edge decision authority to a single operational modality at each vertex in the path from an origin to a destination in a traffic network in order to minimize total network travel time, while utilizing only observed routing decisions of the vehicle. Experimental results demonstrate that TACTS achieves near-optimal performance across varying congestion levels and traffic intensities, generally outperforming state-of-the-art alternatives while maintaining computational efficiency suitable for real-time deployment. These results corroborate the theoretical analysis on the performance bounds of the TACTS algorithm, and suggest that trust-aware control arbitration represents a promising direction for next-generation traffic flow management to leverage changing ground vehicle trends. 

Future work will develop an augmented TACTS algorithm accounting for a traffic network comprised entirely of vehicles with multiple operational modalities, where each vehicle must make an edge decision at each step through the network.

\section*{Acknowledgment}
This work is supported by NSF grant “Satisfaction and Risk-aware Dynamic Resource Orchestration in Public Safety Systems" (SOTERIA) under award \#ECCS-2319995.

\pagebreak

\bibliographystyle{plain}
\bibliography{references}

@article{ait2021quantifying,
  title={Quantifying responses to changes in the jurisdiction of a congestion charge: A study of the London western extension},
  author={Ait Bihi Ouali, Laila and Musuuga, Davis and Graham, Daniel J},
  journal={PLoS One},
  volume={16},
  number={7},
  pages={e0253881},
  year={2021},
  publisher={Public Library of Science San Francisco, CA USA}
}

@article{arnott1991does,
  title={Does providing information to drivers reduce traffic congestion?},
  author={Arnott, Richard and De Palma, Andre and Lindsey, Robin},
  journal={Transportation Research Part A: General},
  volume={25},
  number={5},
  pages={309--318},
  year={1991},
  publisher={Elsevier}
}

@article{bang2022combined,
  title={Combined optimal routing and coordination of connected and automated vehicles},
  author={Bang, Heeseung and Chalaki, Behdad and Malikopoulos, Andreas A},
  journal={IEEE Control Systems Letters},
  volume={6},
  pages={2749--2754},
  year={2022},
  publisher={IEEE}
}

@manual{manual1964bureau,
  title={Traffic Assignment Manual},
  organization={U.S. Department of Commerce, Bureau of Public Roads, Office of Planning, Urban Planning Devision},
  year={1964}
}

@article{bonifaci2010stackelberg,
  title={Stackelberg routing in arbitrary networks},
  author={Bonifaci, Vincenzo and Harks, Tobias and Sch{\"a}fer, Guido},
  journal={Mathematics of Operations Research},
  volume={35},
  number={2},
  pages={330--346},
  year={2010},
  publisher={INFORMS}
}

@article{braess1968paradoxon,
  title={{\"U}ber ein Paradoxon aus der Verkehrsplanung},
  author={Braess, Dietrich},
  journal={Unternehmensforschung},
  volume={12},
  number={1},
  pages={258--268},
  year={1968},
  publisher={Springer}
}

@inproceedings{brown2024tasr,
  title={Tasr: A novel trust-aware stackelberg routing algorithm to mitigate traffic congestion},
  author={Brown, Doris E. M. and Nadendla, Venkata Sriram Siddhardh and Das, Sajal K},
  booktitle={2024 IEEE International Conference on Smart Computing (SMARTCOMP)},
  pages={150--157},
  year={2024},
  organization={IEEE}
}

@INPROCEEDINGS{brown2025iterative,
  author={Brown, Doris E. M. and Nadendla, Venkata Sriram Siddhardh and Das, Sajal K.},
  booktitle={2025 IEEE International Conference on Smart Computing (SMARTCOMP)}, 
  title={Iterative Recommendations Based on Monte Carlo Sampling and Trust Estimation in Multi-Stage Vehicular Traffic Routing Games}, 
  year={2025},
  pages={66-73},
  doi={10.1109/SMARTCOMP65954.2025.00058}}

@inproceedings{craig2021should,
  title={Should self-driving cars mimic human driving behaviors?},
  author={Craig, Jamie and Nojoumian, Mehrdad},
  booktitle={International Conference on Human-Computer Interaction},
  pages={213--225},
  year={2021},
  organization={Springer}
}

@inproceedings{das2017reducing,
  title={Reducing congestion through information design},
  author={Das, Sanmay and Kamenica, Emir and Mirka, Renee},
  booktitle={2017 55th annual allerton conference on communication, control, and computing (allerton)},
  pages={1279--1284},
  year={2017},
  organization={IEEE}
}

@article{gambetta2000can,
  title={Can We Trust Trust},
  author={Gambetta, D},
  journal={Trust: Making and Breaking Cooperative Relation, Electronic edition},
  year={2000}
}

@article{gonzalez2023reducing,
  title={Reducing urban traffic congestion via charging price},
  author={Gonz{\'a}lez-Aliste, Pablo and Derpich, Iv{\'a}n and L{\'o}pez, Mario},
  journal={Sustainability},
  volume={15},
  number={3},
  pages={2086},
  year={2023},
  publisher={MDPI}
}

@article{gu2018congestion,
  title={Congestion pricing practices and public acceptance: A review of evidence},
  author={Gu, Ziyuan and Liu, Zhiyuan and Cheng, Qixiu and Saberi, Meead},
  journal={Case Studies on Transport Policy},
  volume={6},
  number={1},
  pages={94--101},
  year={2018},
  publisher={Elsevier}
}

@article{harrington2001overcoming,
  title={Overcoming public aversion to congestion pricing},
  author={Harrington, Winston and Krupnick, Alan J and Alberini, Anna},
  journal={Transportation Research Part A: Policy and Practice},
  volume={35},
  number={2},
  pages={87--105},
  year={2001},
  publisher={Elsevier}
}

@article{hart2000simple,
  title={A simple adaptive procedure leading to correlated equilibrium},
  author={Hart, Sergiu and Mas-Colell, Andreu},
  journal={Econometrica},
  volume={68},
  number={5},
  pages={1127--1150},
  year={2000},
  publisher={Wiley Online Library}
}

@article{ho2023adaptive,
  title={Adaptive road shoulder traffic control with reinforcement learning approach},
  author={Ho, Yao-Hua and Cheng, Tung-Chun},
  journal={Neural Computing and Applications},
  pages={1--17},
  year={2023},
  publisher={Springer}
}

@inproceedings{isa2014review,
  title={A review on recent traffic congestion relief approaches},
  author={Isa, Norulhidayah and Yusoff, Marina and Mohamed, Azlinah},
  booktitle={2014 4th international conference on artificial intelligence with applications in engineering and technology},
  pages={121--126},
  year={2014},
  organization={IEEE}
}

@article{kashmiri2024routing,
  title={Routing of multi-modal autonomous vehicles for system optimal flows and average travel cost equilibrium over time},
  author={Kashmiri, Faizan Ahmad and Lo, Hong K},
  journal={Transportation research part C: emerging technologies},
  volume={159},
  pages={104483},
  year={2024},
  publisher={Elsevier}
}

@inproceedings{kolarich2022stackelberg,
  title={Stackelberg routing of autonomous cars in mixed-autonomy traffic networks},
  author={Kolarich, Maxwell and Mehr, Negar},
  booktitle={2022 American Control Conference (ACC)},
  pages={4654--4661},
  year={2022},
  organization={IEEE}
}

@article{korilis1997achieving,
  title={Achieving Network Optima Using Stackelberg Routing Strategies},
  author={Korilis, Yannis A and Lazar, Aurel A and Orda, Ariel},
  journal={IEEE/ACM Transactions on Networking},
  volume={5},
  number={1},
  pages={161},
  year={1997}
}

@article{kormos2021cities,
  title={How cities can apply behavioral science to promote public transportation use},
  author={Kormos, Christine and Sussman, Reuven and Rosenberg, Bracha},
  journal={Behavioral Science \& Policy},
  volume={7},
  number={1},
  pages={95--115},
  year={2021},
  publisher={SAGE Publications Sage CA: Los Angeles, CA}
}

@article{krichene2016social,
  title={On social optimal routing under selfish learning},
  author={Krichene, Walid and Castillo, Milena Suarez and Bayen, Alexandre},
  journal={IEEE transactions on control of network systems},
  volume={5},
  number={1},
  pages={479--488},
  year={2016},
  publisher={IEEE}
}

@inproceedings{letchford2009learning,
  title={Learning and approximating the optimal strategy to commit to},
  author={Letchford, Joshua and Conitzer, Vincent and Munagala, Kamesh},
  booktitle={International symposium on algorithmic game theory},
  pages={250--262},
  year={2009},
  organization={Springer}
}

@article{li2023dynamic,
  title={Dynamic lane reversal strategy in intelligent transportation systems in smart cities},
  author={Li, Wenting and Li, Jianqing and Han, Di},
  journal={Sensors},
  volume={23},
  number={17},
  pages={7402},
  year={2023},
  publisher={MDPI}
}

@article{liu2022analysis,
  title={Analysis of the conflict between car commuter’s route choice habitual behavior and traffic information search behavior},
  author={Liu, Kai},
  journal={Sensors},
  volume={22},
  number={12},
  pages={4382},
  year={2022},
  publisher={MDPI}
}

@manual{mobilus2021sae,
  title={{J3016\_202104} - {T}axonomy and Definitions for Terms Related to Driving Automation Systems for On-Road Motor Vehicles},
  organization={{SAE} International},
  year={2021}
}

@article{mostafizi2021decentralized,
  title={A decentralized and coordinated routing algorithm for connected and autonomous vehicles},
  author={Mostafizi, Alireza and Koll, Charles and Wang, Haizhong},
  journal={IEEE Transactions on Intelligent Transportation Systems},
  volume={23},
  number={8},
  pages={11505--11517},
  year={2021},
  publisher={IEEE}
}

@article{moylan2016observed,
  title={Observed and simulated traffic impacts from the 2013 Bay Area Rapid Transit strike},
  author={Moylan, Emily and Foti, Fletcher and Skabardonis, Alexander},
  journal={Transportation Planning and Technology},
  volume={39},
  number={2},
  pages={162--179},
  year={2016},
  publisher={Taylor \& Francis}
}

@article{nguyen2017local,
  title={Local and system-wide traffic effects of urban road-rail level crossings: A new 4 estimation technique},
  author={Nguyen-Phuoc, Duy Q and Currie, Graham and De Gruyter, Chris and Young, William},
  journal={Journal of transport geography},
  volume={60},
  pages={89--97},
  year={2017},
  publisher={Elsevier}
}

@article{pei2023self,
  title={Self-organized routing for autonomous vehicles via deep reinforcement learning},
  author={Pei, Huaxin and Zhang, Jiawei and Zhang, Yi and Xu, Huile and Li, Li},
  journal={IEEE Transactions on Vehicular Technology},
  volume={73},
  number={1},
  pages={426--437},
  year={2023},
  publisher={IEEE}
}

@inproceedings{roughgarden2001stackelberg,
  title={Stackelberg scheduling strategies},
  author={Roughgarden, Tim},
  booktitle={Proceedings of the thirty-third annual ACM symposium on Theory of computing},
  pages={104--113},
  year={2001}
}

@article{sourav2024work,
  title={Work-in-Progress: Traded Control Transfer for Managing Real-Time Sensor Uncertainties in Autonomous Vehicle},
  author={Sourav, Md Sakib Galib and Cheng, Liang},
  journal={arXiv preprint arXiv:2410.06345},
  year={2024}
}

@inproceedings{tavafoghi2017informational,
  title={Informational incentives for congestion games},
  author={Tavafoghi, Hamidreza and Teneketzis, Demosthenis},
  booktitle={2017 55th Annual Allerton Conference on Communication, Control, and Computing (Allerton)},
  pages={1285--1292},
  year={2017},
  organization={IEEE}
}

@manual{tesla2025model3_navigation,
  title        = {Tesla Model 3 Owner’s Manual – Navigation / Selecting Alternate Route},
  author       = {Tesla, Inc.},
  year         = {2025},
  url          = {https://www.tesla.com/ownersmanual/model3/en_is/GUID-01F1A582-99D1-4933-B5FB-B2F0203FFE6F.html},
}

@article{toso2023impact,
  title={Impact on traffic of delayed information in navigation systems},
  author={Toso, Tommaso and Kibangou, Alain Y and Frasca, Paolo},
  journal={IEEE Control Systems Letters},
  volume={7},
  pages={1500--1505},
  year={2023},
  publisher={IEEE}
}

@article{vickery1969congestion,
  title={Congestion theory and transport investment},
  author={Vickery, William},
  journal={American Economic Review},
  volume={59},
  number={2},
  pages={251--60},
  year={1969}
}

@inproceedings{wang2018automation,
  title={Automation Reliability and Trust: A Bayesian Inference Approach},
  author={Wang, Chenlan and Zhang, Chongjie and Yang, X Jessie},
  booktitle={Proceedings of the Human Factors and Ergonomics Society Annual Meeting},
  volume={62},
  pages={202--206},
  year={2018},
  organization={SAGE Publications Sage CA: Los Angeles, CA}
}

@article{wang2020empirical,
  title={Empirical study of effect of dynamic travel time information on driver route choice behavior},
  author={Wang, Jinghui and Rakha, Hesham},
  journal={Sensors},
  volume={20},
  number={11},
  pages={3257},
  year={2020},
  publisher={MDPI}
}

@article{wang2022computational,
  title={Computational Model of Robot Trust in Human Co-Worker for Physical Human-Robot Collaboration},
  author={Wang, Qiao and Liu, Dikai and Carmichael, Marc G and Aldini, Stefano and Lin, Chin-Teng},
  journal={IEEE Robotics and Automation Letters},
  volume={7},
  number={2},
  pages={3146--3153},
  year={2022},
  publisher={IEEE}
}

@misc{waymo_user_route_selection_2023,
  title       = {User‑controlled route selection for autonomous vehicles},
  author      = {Waymo LLC},
  year        = {2023},
  number      = {US 2023/0391363 A1},
  url         = {https://www.freepatentsonline.com/y2023/0391363.html},
}

\end{document}